\documentclass[epj]{svjour}

\usepackage{graphics}
\usepackage{bbm}
\usepackage{amsmath}
\usepackage{hyperref}
\newcommand{\op}{\boldsymbol}

\begin{document}

\title{Quantum Coherence and Path-Distinguishability of Two Entangled Particles}
\author{Misba Afrin\inst{1,2} \and Tabish Qureshi\inst{2}}
\institute{Department of Physics, Jamia Millia Islamia, New Delhi,
\email{me.misba@gmail.com}
\and
Centre for Theoretical Physics, Jamia Millia Islamia, New Delhi,
\email{tabish@ctp-jamia.res.in}}


\abstract{
An interference experiment with entangled particles is theoretically analyzed,
where one of the entangled pair (particle 1) goes through a multi-slit
before being detected at a fixed detector. In addition, one introduces
a mechanism for finding out which of the n slits did particle 1 go through.
The other particle of the entangled
pair (particle 2) goes in a different direction, and is detected at a
variable, spatially separated location. In coincident counting, particle 2
shows n-slit interference.  It is shown that the normalized
quantum coherence of particle 2, $\mathcal{C}_2$, and the
path-distinguishability of particle 1, $\mathcal{D}_{Q1}$, are bounded
by an inequality $\mathcal{D}_{Q1} + \mathcal{C}_2 \le 1$. 
This is a kind of {\em nonlocal} duality relation, which connects
the path distinguishability of one particle to the quantum coherence
of the other. 
}

\maketitle

\section{Introduction}

Quantum coherence is a topic which has been under intense study in recent
years. Coherence has always been an important issue in quantum 
optics \cite{mandelcoherence}, but with the advent of quantum information
and computation there was a need for a rigorous quantitative measure of
quantum coherence. Recently a measure of coherence was introduced,
which is basically the sum of the absolute values of the off-diagonal elements
of the density matrix of a system in a particular basis, i.e.,
$\sum_{i\neq j} |\rho_{ij}|$, 
with $\rho_{i,j}=\langle i | \rho | j \rangle$ \cite{coherence}.
This measure can be normalized for finite dimensional Hilbert space, 
to give the following definition of coherence \cite{nduality}
\begin{equation}
{\mathcal C} = {1\over n-1}\sum_{i\neq j}|\langle i |\rho| j \rangle| ,
\label{coherence}
\end{equation}
where $n$ is the dimensionality of the Hilbert space. 
This measure has proven to be particularly useful in quantifying 
wave-particle duality in n-path interference \cite{nduality,bagan,nslit}.

Neils Bohr's assertion that the wave and particle natures are mutually
exclusive, came to be known as the principle of complementarity,
or more popularly as wave-particle duality \cite{bohr}. 
This principle has stood its ground despite criticism and attacks over the
years. It has also been given a quantitative meaning by a
bound on the extent to which the two natures could be simultaneously observed
\cite{greenberger,englert}. 
More recently Bohr's complementarity principle has been generalized to
n-path interference. The principle can be quantitatively stated, for a
n-slit interference where one tries to extract information on which slit
the particle went through, by the following duality relation
\begin{equation}
\mathcal{D}_Q + \mathcal{C} \le 1,
\label{nduality}
\end{equation}
where $\mathcal{C}$ is the coherence defined by (\ref{coherence}), and
$\mathcal{D}_Q$ is a measure of how much path information can be obtained
from the particle (to be formally defined later). Another definition of
path-distinguishability leads to a different form of duality relation
\cite{nslit}.

Needless to say, when 
we talk of path-distinguishability, we talk of the path 
knowledge of the same particle which contributes to the interference
pattern. Thus the relation (\ref{nduality}) is {\em local}, which might
look like stating the obvious. However, in this study 
we propose and theoretically analyze an experiment involving pairs
of entangled particles in which we relate the path information
of one particle to the coherence of the other.

\section{N-slit ghost interference}

We start by generalizing the well known ghost-interference
experiment carried out by Strekalov et al.\cite{ghostexpt}. In the original
experiment, pairs of entangled photons generated from a
spontaneous parametric down conversion (SPDC) source, are separated by
a polarizing beam-splitter. Photon 1 passes through a double-slit before
being registered in a {\em fixed}
detector D2. Photon 2 travels undisturbed before being detected by 
the scanning detector D2. The detectors D1 and D2 are connected to a 
coincidence counter. The detector D2 for photon 2, when counted in coincidence
with the fixed detector D1, shows a two-slit interference pattern,
although photon 2 does not pass through any double-slit.
This interference, called ghost interference, has been understood to be a
consequence of entanglement, and generated lot of research activity
\cite{ghostimaging,rubin,zhai,jie,zeil2,pravatq,twocolor,sheebatq,ghostunder}.
\begin{figure}
\centerline{\resizebox{8.5cm}{!}{\includegraphics{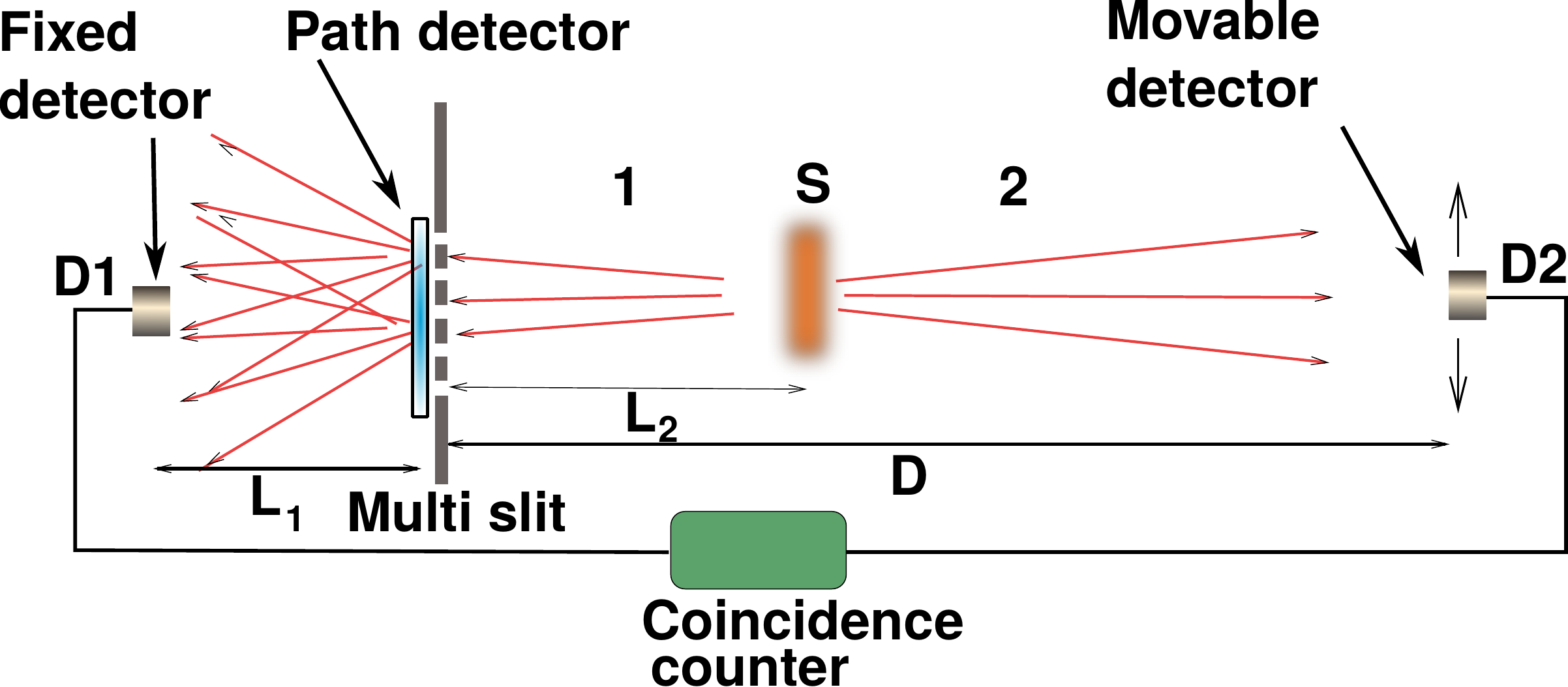}}}
\caption{A modified ghost-interference experiment with entangled pairs of
particles. A which-path detector can detect which of the n slits
particle 1 went through. Detector D1 is fixed at $z=0$, but D2 is free to
move along the z-axis.}
\label{ghostmod}
\end{figure}

We look at a modified ghost interference experiment, shown in
Fig.\ref{ghostmod}, where the double-slit is replaced by a n-slit.
A n-slit ghost interference experiment has actually been performed by
Zeilinger's group \cite{zeil1}. However, here we will consider another
modification of it.
Entangled particle pairs, which could also more generally be something other than
photons, emerge from a source S. 
Particle 1 passes through a n-slit and also interacts with
a path-detector. We do not assume any form of the path-detector,
but just assume that it is an n-state system with states $|d_1\rangle,
|d_2\rangle, |d_3\rangle,\dots, |d_n\rangle$,
which get entangled with the n paths of particle 1.
This entanglement is bound to happen if the path-detector acquires
the relevant information about which slit particle 1 went through.
Particle 1 then travels further and is detected by a {\em fixed} detector D1.
Particle 2 travels undisturbed to detector D2. Since the two particles have
to be counted in coincidence,
the paths traveled by both the particles, before reaching their respective
detectors, are assumed to be equal. Without the path-detector, our experiment
would just be a n-slit generalization of the original two-slit ghost
interference experiment \cite{ghostexpt}. It is expected to show a
n-slit ghost interference for particle 2. By introducing a path-detector,
we are probing if acquiring path
information about particle 1, has any effect on the interference shown by
particle 2. A somewhat related question, namely, how the wave-particle
duality of one particle is affected if it is entangled with another particle,
has already been explored \cite{bu}.

We assume $|d_1\rangle,\dots, |d_n\rangle$ to be normalized, but not
necessarily
orthogonal. The ultimate limit to the knowledge we can acquire as to
which slit particle 1 went through, is set by how distinct the states
$|d_1\rangle,\dots, |d_n\rangle$  are. If $|d_1\rangle,\dots, |d_n\rangle$ 
are orthogonal, we can {\em in principle} know with certainty which slit the
particle went through. Of course in general, $|d_1\rangle,\dots, |d_n\rangle$
may not be all orthogonal to each other. In such a situation, one is 
left with the problem of unambiguously telling which of the states
$|d_1\rangle,\dots, |d_n\rangle$, is the given unknown path-detector state
which the path-detection process throws up. The best bet to answer this
question is using unambiguous quantum state discrimination (UQSD)
\cite{uqsd,dieks,peres,jaeger2,bergou}. 

Using UQSD, a path-distinguishability ${\mathcal D}_Q$ can then be defined as
the upper bound to the probability with which the states
$|d_1\rangle,\dots, |d_n\rangle$ can be distinguished
without any error \cite{3slit,nduality}. If the state $|d_k\rangle$ occurs
with a probability $|c_k|^2$, the path distinguishability has the following
form \cite{3slit,nduality}
\begin{equation}
{\mathcal D}_Q \equiv 1 - {1\over n-1}\sum_{i\neq j} |c_i||c_j| |\langle d_i|d_j
\rangle|.
\label{D}
\end{equation}
The path-distinguishability can take values between 0 and 1. 

\section{General analysis}

Next we formulate the n-slit ghost interference, with a path-detector, in a most
general way. We assume that the
particles travel in opposite directions along the x-axis. The entanglement
is in the z-direction. The entangled two-particle state at the source,
at time $t=0$, is given by $|\Psi(0)\rangle$.
We assume that after traveling for a time $t_0$, particle 1 reaches the 
n-slit ($vt_0 = L_2$), and particle 2 travels a distance $L_2$ towards
detector D2. The two particle state is now given by $|\Psi(t_0)\rangle$.

We take into account the effect of the n-slit on the 
entangled state as follows. We assume that the n-slit allows
the portions of the wave-function in front of the slits to pass through,
and blocks the other portions. We assume that what
emerge from the n-slit are localized states, whose
width is approximately the width of a slit. The  states of particle 1,
which pass through the slits $1, 2,\dots, n$, are denoted by
$|\phi_1\rangle, |\phi_2\rangle,\dots, |\phi_n\rangle$, respectively.
The state representing the situation in which particle 1 gets blocked is, say,
$|\chi\rangle$. These n+1 states are obviously orthogonal, because they
represent mutually exclusive possibilities, and form a complete set because
they exhaust all the possibilities for particle 1. Thus the
entangled two-particle state can be expanded in terms of these.
We can thus write:
\begin{eqnarray}
|\Psi(t_0)\rangle &=& \left(\sum_{k=1}^n |\phi_k\rangle\langle\phi_k|
+|\chi\rangle\langle\chi|\right)\otimes \mathbbm{1}_2
 |\Psi(t_0)\rangle \nonumber\\
&=& \sum_{k=1}^n |\phi_k\rangle\langle\phi_k|\Psi(t_0)\rangle 
+|\chi\rangle\langle\chi|\Psi(t_0)\rangle ,
\end{eqnarray}
where $\mathbbm{1}_2$ is the unit operator for the space of particle 2.
Now, since we are only interested in those instances (through coincidence
counting) where particle 1
does pass through the multi-slit, and does not get blocked, the 
$|\chi\rangle$ dependent term can be discarded, and one is left with the
two-particle state
\begin{eqnarray}
|\Psi(t_0)\rangle &=& 
\sum_{k=1}^n |\phi_k\rangle\langle\phi_k|\Psi(t_0)\rangle ,
\end{eqnarray}
which needs to be normalized again. A typical term
$\langle\phi_k|\Psi(t_0)\rangle$ represents an unnormalized
state of particle 2. Normalizing it will essentially throw up a constant
specific to that state:
\begin{eqnarray}
\langle\phi_k|\Psi(t_0)\rangle = c_k |\psi_k\rangle ,
\end{eqnarray}
where $|\psi_k\rangle$ are normalized states of particle 2.
Particle 1, emerging from the n-slit, interacts with a path-detector which
is initially in the state $|d_0\rangle$. The normalized state of the two
particles, plus the path detector, is given by
\begin{eqnarray}
|\Psi(t_0)\rangle &=& 
\left(\sum_{k=1}^n c_k |\phi_k\rangle|\psi_k\rangle \right)|d_0\rangle.
\end{eqnarray}
In addition, the states of particle 1 get entangled with the n
states of the which-path detector $|d_1\rangle,\dots, |d_n\rangle$.
So, the state we get after particle 1 crosses the n-slit is:
\begin{eqnarray}
|\Psi\rangle &=& 
\sum_{k=1}^n c_k |\phi_k\rangle|\psi_k\rangle|d_k\rangle.
\label{Psif}
\end{eqnarray}

Now, if one is only interested in the path detector, for all practical
purposes it is in a mixed state with $|d_k\rangle$ occurring with a
probability $|c_k|^2$, as is obvious from (\ref{Psif}).
Path distinguishability of particle 1 can then be simply written down,
using (\ref{D}), as
\begin{equation}
{\mathcal D}_{Q1} = 1 - {1\over n-1}\sum_{i\neq j} |c_i||c_j| |\langle d_i|d_j
\rangle|.
\label{D1}
\end{equation}
The above equation quantifies the amount of path knowledge about particle
1 one can obtain, given the path-detector states $\{|d_k\rangle\}$.

After interacting with the path-detector, particle 1 travels to the fixed
detector D1. Particle 2 continues its travel undisturbed to reach D2,
and should give rise to interference. The state of the two particle just before
hitting the detectors is given by
\begin{eqnarray}
|\Psi_s\rangle &=& 
\sum_{k=1}^n c_k\op{U_1}|\phi_k\rangle\op{U}_2  |\psi_k\rangle|d_k\rangle,
\label{Psis}
\end{eqnarray}
where $\op{U_1}, \op{U_2}$ represent the time evolution operators for
particle 1 and 2, respectively, from time $t_0$ to
the times the particles hit the detectors.
Interference is a signature of the wave nature. It has been
argued before that the wave nature of a quanton, in an interference experiment,
can be be quantified by its coherence $\mathcal{C}$, defined by
(\ref{coherence}) \cite{nduality}. It has also been demonstrated that it is possible to
actually measure $\mathcal{C}$ in an interference experiment \cite{tania}.
For particle 2, the coherence $\mathcal{C}$ which will quantify its wave
nature, is given by
\begin{equation}
\mathcal{C}_2 = {1\over n-1}\sum_{i\neq j} |\langle\psi_i|\op{U_2^\dag}\rho_r\op{U_2}|\psi_j\rangle|,
\label{coherence2}
\end{equation}
where $\rho_r$ is the reduced density matrix of particle 2, obtained after
tracing over the states of particle and the path-detector. In writing the
above, we have tacitly assumed that $\{|\psi_k\rangle\}$ form an orthonormal
set. That is actually an assumption, and will be true only if the 
entanglement between the two particles is good. For example, if
$|\Psi(t_0)\rangle$ is the so-called EPR state \cite{epr}, all
$|\psi_k\rangle$s will be orthogonal to each other. If the $|\psi_k\rangle$s
are not mutually orthogonal, $\mathcal{C}_2$ cannot be calculated using
(\ref{coherence2}), although interference may still arise.
Let us assume that this condition is satisfied, and $|\psi_k\rangle$s are
mutually orthogonal, and go ahead with calculating $\mathcal{C}_2$.

Since $\{|\phi_k\rangle\}$ form an orthonormal set,
$\{\op{U_1}|\phi_k\rangle\}$ will also form an orthonormal set.
If one evaluates the reduced density operator for particle 2 by using
$|\Psi_s\rangle$ given by (\ref{Psis}), and employing the basis
$\{\op{U_1}|\phi_k\rangle\}$ for particle 1, one gets the following result:
\begin{eqnarray}
\rho_r &=& Tr_d\left[ \sum_k \langle\phi_k|\op{U_1^\dag} \left(|\Psi_s\rangle\langle\Psi_s|\right)
\op{U_1}|\phi_k\rangle\right] \nonumber\\
 &=&  \sum_{k,k',k''} c_{k'}c_{k''}^*\langle\phi_k|\op{U_1^\dag} \left(\op{U_1}|\phi_{k'}\rangle\op{U}_2  |\psi_{k'}\rangle\right.\nonumber\\
&&\left. \langle\psi_{k''}|\op{U_2^\dag}\langle\phi_{k''}|\op{U_1^\dag}\right)
\op{U_1}|\phi_k\rangle \langle d_{k''}|d_{k'}\rangle \nonumber\\
&=& \sum_k |c_k|^2\op{U_2}|\psi_k\rangle\langle\psi_k|\op{U_2^\dag} ,
\end{eqnarray}
where $Tr_d$ represents a trace over the path-detectors states. The above
is clearly diagonal in the basis $\{\op{U_2}|\psi_k\rangle\}$, and consequently 
yields $\mathcal{C}_2 = 0$. This means no interference. The reason for this
apparently negative result is that the interference in particle 2 is not
first order. It only occurs when the particles are detected in coincidence
with a fixed D1. Let us assume that particle 2 is counted only when 
particle 1 is found in a state $|z_0\rangle$, which may be a state localized 
in position. Then the reduced density operator for particle 2 
is given by
\begin{eqnarray}
\rho_r &=& \frac{Tr_d\left[ \langle z_0|\Psi_s\rangle\langle\Psi_s|z_0\rangle \right]}
{Tr\left[ \langle z_0|\Psi_s\rangle\langle\Psi_s|z_0\rangle \right]}
\nonumber\\
&=& \frac{\sum_{j,k} c_jc_k^*\langle z_0|\op{U_1}|\phi_j\rangle\langle\phi_k|
\op{U_1^\dag}|z_0\rangle
\langle d_j|d_k\rangle|\psi_j\rangle\langle\psi_k|}
{\sum_{k} |c_k|^2|\langle z_0|\op{U_1}|\phi_k\rangle|^2 } .\nonumber\\
\label{rhor}
\end{eqnarray}
Using (\ref{rhor}) and (\ref{coherence2}), $\mathcal{C}_2$ can be easily
worked out to give
\begin{eqnarray}
\mathcal{C}_{2} &=& {1\over n-1}\frac{\sum_{i\neq j} |c_i||c_j|~|\langle d_i|d_j
\rangle|~|\langle z_0|\op{U_1}|\phi_i\rangle||\langle z_0|\op{U_1}|\phi_j\rangle|}
{\sum_{k} |c_k|^2|\langle z_0|\op{U_1}|\phi_k\rangle|^2 }\nonumber\\
 &\le& {1\over n-1}\sum_{i\neq j} |c_i||c_j| |\langle d_i|d_j\rangle|,
\end{eqnarray}
where the inequality is saturated when all
$|\langle z_0|\op{U_1}|\phi_k\rangle|$ are equal.
The above relation, together with (\ref{D1}), results in the following
inequality
\begin{equation}
\mathcal{D}_{Q1} + \mathcal{C}_{2} \le 1.
\label{gduality0}
\end{equation}
This relation puts a bound on the coherence of particle 2, and the
amount of path information which can be extracted for particle 1. This is a 
completely nonlocal effect, and is a consequence of the entanglement between
the two particles.

\section{Wave-packet analysis}

The analysis in the preceding section serves to reveal the general
nature of ghost interference, and provides bounds on the wave and particle
natures of the two entangled particles. However, it's applicability is
restricted to the situation where $\{|\psi_k\rangle\}$ are mutually
orthogonal. If that condition is not satisfied, coherence cannot be 
calculated by the method presented in that analysis. However, there is
another way in which coherence can be calculated, without using knowledge
of a basis. That is possible by analyzing the interference pattern \cite{tania}.
But for that we need the functional form of the interference pattern, and
the analysis should involve functional form of the wave-functions involved.
In the following, we carry out a wave-packet analysis of the n-slit 
ghost interference, and evaluate coherence by this other method.

First we need a functional form of the two-particle entangled state
$|\Psi(0)\rangle$.  Momentum-entangled particles can be described very
nicely using the {\em generalized EPR state} \cite{tqajp}
\begin{equation}
\Psi(z_1,z_2) = C\!\int_{-\infty}^\infty dk~
e^{-k^2/4\sigma^2}e^{-ikz_2} e^{i kz_1}
e^{-{(z_1+z_2)^2\over 4\Omega^2}}, \label{state}
\end{equation}
where $C$ is a normalization constant, and $\sigma,\Omega$ are certain
parameters. In the limit $\sigma,\Omega\to\infty$ the state (\ref{state})
reduces to the so-called EPR state introduced by Einstein, Podolsky and
Rosen \cite{epr}.
After performing the integration over $p$, (\ref{state}) reduces to
\begin{equation}
\Psi(z_1,z_2) = \sqrt{ {\sigma\over \pi\Omega}}
 e^{-(z_1-z_2)^2\sigma^2} e^{-(z_1+z_2)^2/4\Omega^2} .
\label{psi0}
\end{equation}
It is straightforward to show that $\Omega$ and $\sigma$ quantify the position
and momentum spread of the particles in the z-direction.
We would like to reemphasize that the two particles are assumed to be
moving along the x-axis, in opposite directions, and the slits are in
the y-z plane, each slit being parallel to the y-axis, placed at a 
different z position. The dynamics of the particle along the x-axis is
uninteresting, and only serves to transport the particle from the source
to the detectors. Consequently we will ignore this dynamics and will just
assume that particle 1, moving with an average momentum $p_0$, reaches
the n-slit at a time $t_0$, after traveling a distance $L_2$. A deBroglie
wavelength can be associated with the particle, $\lambda=h/p_0$.

The state of the entangled system, after this time evolution, can be
calculated using the Hamiltonian governing the time evolution, given by
$\hat{H} = {p_1^2\over 2m} +{p_2^2\over 2m}$.
After a time $t_0$, (\ref{psi0}) assumes the form
\begin{eqnarray}
\Psi(z_1,z_2,t_0) = C_{t_0}
\exp\left[\tfrac{-(z_1-z_2)^2}{ {1\over\sigma^2} + {4i\hbar t_0\over m}}
\right]
\exp\left[\tfrac{-(z_1+z_2)^2}{ \left(4\Omega^2  +
{i\hbar t_0\over m}\right)} \right],
\label{Psit}
\end{eqnarray}
where $C_{t_0}=\left({\pi}(\Omega+{i\hbar t_0\over 4m\Omega})
(1/\sigma + {4i\hbar t_0\over m/\sigma})\right)^{-1/2}$.

In the following, we assume that $|\phi_k\rangle$ are Gaussian wave-packets:
\begin{eqnarray}
\phi_k(z_1) &=& {1\over(\pi/2)^{1/4}\sqrt{\epsilon}} e^{-(z_1-kz_0)^2/\epsilon^2},
\label{phik}
\end{eqnarray}
where $k z_0$ is the z-position of the k'th slit, and $\epsilon$
its width. Thus, the distance between j'th and k'th slits is
$(j-k)z_0$.

Using (\ref{phik}) and (\ref{Psit}), the wave-function for $|\psi_k\rangle$
can be calculated, which, after normalization, has the form
\begin{equation}
\psi_k(z_2) = C_2 e^{-{(z_2 - kz_0')^2 \over \Gamma}} ,
\end{equation}
where $C_2 = (2/\pi)^{1/4}(\sqrt{\Gamma_R} + {i\Gamma_I\over\sqrt{\Gamma_R}})^{-1/2}$,
\begin{equation}
z_0' = {z_0 \over {4\Omega^2\sigma^2+1\over
4\Omega^2\sigma^2-1} + {4\epsilon^2 \over 4\Omega^2-1/\sigma^2}},
\end{equation}
and
\begin{equation}
\Gamma = \frac{{1\over\sigma^2}+\left(1+\frac{1}{4\sigma^2\Omega^2}\right)
\left(\epsilon^2 + {2i\hbar t_0\over m}\right) }
{1 + {1\over 4\Omega^2\sigma^2} + {\epsilon^2\over\Omega^2}+{i2\hbar t_0\over \Omega^2m} 
} + {2i\hbar t_0\over m}.
\end{equation}
Thus, the state which emerges from the n-slit, has the following form
\begin{eqnarray}
\Psi(z_1,z_2) &=& C_0\sum_{k-1}^n c_k |d_k\rangle e^{{-(z_1-kz_0)^2\over\epsilon^2}}
e^{{-(z_2 - kz_0')^2 \over \Gamma}+i\theta_k} 
\end{eqnarray}
where $C_0 = (1/\sqrt{\pi\epsilon})(\sqrt{\Gamma_R} +
{i\Gamma_I\over\sqrt{\Gamma_R}})^{-1/2}$, $\Gamma_R, \Gamma_I$ being the real
and imaginary parts of $\Gamma$, respectively, and $c_k^2$ is the probability of
particle 1 to emerge from the k'th slit. The constants $c_k$ and the states
$|d_k\rangle$ are assumed to
be real, as any complex phases can be absorbed in $\theta_k$.
Particles travel for another time $t$ before reaching their respective 
detectors. We assume that the wave-packets travel in the x-direction with
a velocity $v_0$ such that $\lambda=h/mv_0$ is the d'Broglie wavelength.
Using this strategy, we can
write $\hbar (t+2t_0)/m = \lambda D/2\pi$, $\hbar t_0/m = \lambda L_2/2\pi$.
The expression $\lambda D/2\pi$ will also hold for a
photon provided, one uses the wavelength of the photon for
$\lambda$\cite{ghostunder}. 
The state acquires the form
\begin{eqnarray}
\Psi(t)=C_t \sum_{k=1}^n c_k
|d_k\rangle e^{-\tfrac{(z_1-kz_0)^2}{ \epsilon^2+{iL_1\lambda\over\pi}}}
 e^{-\tfrac{(z_2 - kz_0')^2 }{ \Gamma+{iL_1\lambda\over\pi}}}
e^{i\theta_k},
\label{psifinal}
\end{eqnarray}
where 
\begin{equation}
C_t = {1\over \sqrt{\pi}\sqrt{\epsilon+iL_1\lambda/\epsilon\pi}
\sqrt{\sqrt{\Gamma_R}+(\Gamma_I+iL_1\lambda/\pi)/\sqrt{\Gamma_R}}}.
\end{equation}

In order to get simplified results, we 
consider the limit $\Omega \gg \epsilon$ and
$\Omega \gg 1/\sigma$. In this limit
\begin{equation}
\Gamma \approx \gamma + 4i\hbar t_0/m ,
\label{ent-strong}
\end{equation}
where $\gamma = \epsilon^2 + 1/\sigma^2$ and $z_0' \approx z_0$.

Using (\ref{psifinal}), we can now calculate the probability of coincident
detection at D1 and D2.
Assuming that D1 is fixed at $z_1=0$, this probability density is given by
$P(z_2) \equiv |\Psi(0,z_2,t)|^2$, which has the following form
\begin{eqnarray}
P(z_2) &=&|C_t|^2\sum_{k=1}^n c_k^2 e^{-\frac{2k^2z_0^2}{\alpha}}
e^{-\frac{2(z_2-kz_0)^2}{\beta}} + \nonumber\\
 && |C_t|^2\sum_{j\ne k}c_jc_k |\langle d_j|d_k\rangle|e^{-\frac{(k^2+j^2)z_0^2}{\alpha}}
e^{-\frac{(z_2-jz_0)^2}{\beta}} e^{-\frac{(z_2-kz_0)^2}{\beta}} \nonumber\\
&& \cos\left[\tfrac{2\pi(k-j)z_2z_0\lambda D}{ \gamma^4\pi^2+\lambda^2 D^2}
+\tfrac{\pi(k^2-j^2)z_0^2\lambda L_1}{ \epsilon^4\pi^2+\lambda^2 L_1^2}
+\tfrac{\pi(k^2-j^2)z_0^2\lambda D}{ \gamma^4\pi^2+\lambda^2 D^2}\right.
\nonumber\\
&&\left.+ \theta_j-\theta_k\right],
\label{pattern}
\end{eqnarray}
where $\alpha=\epsilon^2+\lambda^2 L_1^2/\pi^2\epsilon^2$ and
$\beta = \gamma^2+\lambda^2 D^2/\pi^2\gamma^2$.
Eqn. (\ref{pattern}) represents a n-slit ghost interference pattern for
particle 2, even though it has not passed through any slit.
If the position of of D2, $z_2$ is on any primary maximum away from the one
at $z_2 = 0$, $kz_0$  is negligible in its comparison. This happens basically
because the Gaussians $e^{-\frac{(z_2-kz_0)^2}{\beta}}$ are very broad because 
$\gamma$ is very small, and as a result $\beta$ very large.
Keeping this in mind,
(\ref{pattern}) can be further simplified to:
\begin{eqnarray}
P(z_2) &=&|C_t|^2 e^{-\frac{2z_2^2}{\beta}}\left(\sum_{k=1}^n c_k^2 e^{-\frac{2k^2z_0^2}{\alpha}}
\right. \nonumber\\
 &&\left. + \sum_{j\ne k}c_jc_k |\langle d_j|d_k\rangle|e^{-\frac{(k^2+j^2)z_0^2}{\alpha}}\right.\nonumber\\
&&\left. \cos\left[\frac{2\pi(k-j)z_2z_0\lambda D}{ \gamma^4\pi^2+\lambda^2 D^2}
+ \theta_j-\theta_k\right]\right).
\label{patterns}
\end{eqnarray}

We can calculate the coherence from the interference formed by particle 2.
It is has been demonstrated earlier that the coherence can be calculated from
a n-slit interference pattern as 
$\mathcal{C} = {1\over n-1}{I_{max}-I_{inc}\over I_{inc}}$, where
$I_{max}$ is the maximum intensity at a primary maximum, and $I_{inc}$
is the intensity at the same position if $\theta_j,\theta_k$ were varying
randomly \cite{tania}. The effect of randomly varying phases
$\theta_j,\theta_k$ on (\ref{patterns}) will be that the cosine term will
become zero.
The quantum coherence for particle 2, from (\ref{patterns}), is given by
\begin{eqnarray}
\mathcal{C}_2 &=& {1\over n-1}\frac{\sum_{j\ne k}c_jc_k |\langle d_j|d_k\rangle|e^{-\frac{(k^2+j^2)z_0^2}{\alpha}}}
{\sum_{k=1}^n c_k^2 e^{-\frac{2k^2z_0^2}{\alpha}}}\nonumber\\
 &\le& {1\over n-1}\sum_{j\ne k}c_jc_k |\langle d_j|d_k\rangle|,
\label{C2w}
\end{eqnarray}
where we have used $\sum_{k=1}^n c_k^2 = 1$. Using the above and (\ref{D1}),
one can write
\begin{equation}
\mathcal{D}_{Q1} + \mathcal{C}_{2} \le 1.
\label{gduality}
\end{equation}
That is the same relation which was derived in (\ref{gduality0}), in
the preceding section, from a more general analysis.

We can consider another limit which is opposite to that described by
(\ref{ent-strong}), namely where the entanglement is weak. This is the 
case where $\Omega \approx 1/\sigma$. Here $\Gamma$ can be approximated by
\begin{equation}
\Gamma \approx \gamma + {2i\hbar t_0\over m} ,
\label{ent-weak}
\end{equation}
where $\gamma = 1/2\sigma^2$. Here $\beta$
($= \gamma^2+\lambda^2 D^2/\pi^2\gamma^2$) is independent of $\epsilon$
and does not grow much with time. Consequently, the Gaussians 
$e^{-\frac{(z_2-kz_0)^2}{\beta}}$ in (\ref{pattern}) are not very broad and
may not overlap strongly with each other. This will lead to a reduced 
value of coherence $\mathcal{C}_2$. The path distinguishability
$\mathcal{D}_{Q1}$ of particle 1, on the other hand, is unaffected by the
degree of entanglement. Thus the inequality (\ref{gduality}) remains far
from saturation if the two particles are weakly entangled.

\section{Conclusion}

In conclusion,
we have theoretically analyzed a modified ghost interference experiment
with n-slits and a path-detector behind the multi-slit. We have shown
that extracting path-knowledge about particle 1, affects the coherence of
particle 2, although they are spatially separated. We have shown that a
kind of non-local wave-particle duality relation applies for such a 
situation. Similar studies have been carried out for two-slit and three-slit
experiments, but using fringe visibility, and not coherence
\cite{gduality2,gduality3}.

A critic might be tempted to infer that particle 1 is actually providing the
path information of particle 2, and the inequality (\ref{gduality}) is
essentially a duality relation for a one particle only. There are
several quantum optics experiments where entanglement has indeed been used
to infer path information of a particle, by looking at its entangled 
partner \cite{scarcelli}.  However, in our case, as particle
2 does not pass through any slits, its path information has no meaning.
Particle 2 only shows an interference, without passing through any slits,
and the path information that is obtained, is only of particle 1.

\section*{Acknowledgments}
Misba Afrin thanks the Centre for Theoretical Physics, Jamia Millia Islamia
for providing the facilities of the Centre during the course of this
work.

\section*{Author contribution statement}
Tabish Qureshi formulated the problem and carried out part of the calculations.
Misba Afrin carried out the major part of the calculations.


\begin{thebibliography}{0}

\bibitem{mandelcoherence} L. Mandel, `Coherence and indistinguishability,"
{\em Opt. Lett.} {\bf 16}, 1882-1883 (1991).

\bibitem{coherence} T. Baumgratz, M. Cramer, M. B. Plenio, {\em Phys. Rev. Lett.
} {\bf 113}, 140401 (2014).

\bibitem{nduality} M.N. Bera, T. Qureshi, M.A. Siddiqui, A.K. Pati, {\em Phys. Rev. A} {\bf 92}, 012118 (2015).

\bibitem{bagan} E. Bagan, J.A. Bergou, S.S. Cottrell, M. Hillery, ``Relations be
tween Coherence and Path Information," {\em Phys. Rev. Lett.} {\bf 116}, 160406 
(2016).

\bibitem{nslit}  T. Qureshi, M.A. Siddiqui, ``Wave-particle duality in N-path interference", {\em Ann. Phys.} {\bf 385}, 598-604 (2017).

\bibitem{bohr} N. Bohr, ``The quantum postulate and the recent development of
atomic theory," Nature (London) 121, 580-591 (1928). 

\bibitem{greenberger} D. M. Greenberger and A. Yasin,
``Simultaneous wave and particle knowledge in a neutron interferometer",
Phys. Lett. A 128, 391 (1988).

\bibitem{englert} B-G. Englert, ``Fringe visibility and which-way information:
an inequality", {\em Phys. Rev. Lett.} {\bf 77}, 2154 (1996).

\bibitem{ghostexpt} D.V. Strekalov, A.V. Sergienko, D.N. Klyshko, Y.H. Shih,
Observation of two-photon ghost interference and diffraction,
{\em Phys. Rev. Lett.} {\bf 74} 3600 (1995).

\bibitem{ghostimaging} M. D'Angelo, Y-H. Kim, S. P. Kulik and Y. Shih,
{\em Phys. Rev. Lett.} {\bf 92}, 233601 (2004).

\bibitem{rubin} S. Thanvanthri and M. H. Rubin,
{\em Phys. Rev. A} {\bf 70}, 063811 (2004).

\bibitem{zhai} Y-H. Zhai, X-H. Chen, D. Zhang, L-A. Wu,
\textit{Phys. Rev. A} {\bf 72}, 043805. (2005).

\bibitem{jie} L. Jie, C. Jing,
\textit{Chinese Phys. Lett.} {\bf 28}, 094203 (2011).

\bibitem{zeil2} J. Kofler, M. Singh, M. Ebner, M. Keller, M. Kotyrba, A. Zeiling
er,
\textit{Phys. Rev. A} {\bf 86}, 032115 (2012).

\bibitem{pravatq} P. Chingangbam, T. Qureshi,
{\em Prog. Theor. Phys.} {\bf 127}, 383-392 (2012).

\bibitem{twocolor} D-S. Ding, Z-Y. Zhou, B-S. Shi, X-B Zou, G-C. Guo,
``Two-color ghost interference," {\em AIP Advances} {\bf 2}, 032177 (2012).

\bibitem{sheebatq} S. Shafaq, T. Qureshi, ``Theoretical analysis of two-color
ghost interference," {\em Eur. Phys. J. D} {\bf 68}, 52 (2014).

\bibitem{ghostunder}  T. Qureshi, P. Chingangbam, S. Shafaq, ``Understanding
ghost interference," {\em Int. J. Quant. Inf.} {\bf 14(3)}, 1640036 (2016).

\bibitem{zeil1} R. Christanell, W. Weinfurter, A. Zeilinger, {\em The
Technical Digest of the European Quantum Electronic Conference},
EQEC'93, Florence, 1993 (unpublished), p. 872.

\bibitem{bu} K. Bu, L. Li, J. Wu, S-M. Fei, ``Duality relation between coherence and path information in the presence of quantum memory," {\em J. Phys. A: Math. Theor.} {\bf 51}, 085304 (2018).

\bibitem{uqsd}  I.D. Ivanovic, {\em Phys. Lett. A} {\bf 123}, 257 (1987).

\bibitem{dieks}  D. Dieks, 
{\em Phys. Lett. A} {\bf 126}, 303 (1988).

\bibitem{peres}  A. Peres, {\em Phys. Lett. A} {\bf 128}, 19 (1988).

\bibitem{jaeger2} G. Jaeger, A. Shimony, {\em Phys. Lett. A} {\bf 197}, 83 (1995).

\bibitem{bergou} J.A. Bergou, U. Herzog, M. Hillery, {\em Lect. Notes Phys.} {\bf 649}, 417-465 (2004).

\bibitem{3slit}  M. A. Siddiqui, T. Qureshi, {\em Prog. Theor. Exp. Phys.} {\bf 2015}, 083A02 (2015).

\bibitem{tania} T. Paul, T. Qureshi, ``Measuring quantum coherence in multi-slit interference,"
 {\em Phys. Rev. A} {\bf 95}, 042110 (2017).

\bibitem{epr} A. Einstein, B. Podolsky, N. Rosen, ``Can quantum-mechanical
description of physical reality be considered complete?",
\textit{Phys. Rev.} {\bf 47} (1935) 777--780.

\bibitem{tqajp} T. Qureshi, ``Understanding Popper's experiment,"
{\em Am. J. Phys.} {\bf 73}, 541-544 (2005).

\bibitem{gduality2} M.A. Siddiqui, T. Qureshi, ``A nonlocal wave–particle duality," {\em Quantum Stud.: Math. Found.} {\bf 3}, 115–122 (2016).

\bibitem{gduality3} M.A. Siddiqui, ``Three-slit ghost interference and non-local duality," {\em Int. J. Quant. Inf.} {\bf 13}, 1550022 (2015).

\bibitem{scarcelli} G. Scarcelli, Y. Zhou, Y. Shih, ``Random delayed-choice
quantum eraser via two-photon imaging,"
{\em Eur. Phys. J. D} {\bf 44}, 167-173 (2007).

\end{thebibliography}
\end{document}